\documentclass[showpacs,aps,amsmath,amssymb,twocolumn]{revtex4}
\usepackage[latin1]{inputenc}
\usepackage{graphicx}
\usepackage{bm}
\usepackage{amsmath}
\usepackage{array}
\usepackage{braket}
\usepackage{microtype}
\usepackage{appendix}
\usepackage{xcolor}

\renewcommand{\baselinestretch}{1.0}
\begin{document}

\renewcommand{\baselinestretch}{1.0}

\title{A Newtonian approach to the study of irreversibility in many-body systems}

\author{Bertúlio de Lima Bernardo$^{1,2}$}

\affiliation{$^{1}$Departamento de Física, Universidade Federal da Paraíba, 58051-900 João Pessoa, PB, Brazil\\
$^{2}$Departamento de Física, Universidade Federal de Campina Grande, Caixa Postal 10071, 58109-970 Campina Grande-PB, Brazil}

\email{bertulio.fisica@gmail.com}

\begin{abstract}

Irreversibility remains one of the least understood concepts in physics. One of the main reasons is the fact that the fundamental laws of classical and quantum physics are time symmetric, whereas macroscopic processes evolve in a preferred time direction. This long-standing conflict of ideas goes to the heart of the so-called Loschmidt's paradox. Here, we address this dichotomy from a Newtonian perspective of the dynamics of the center of mass of a many-body system, which was found to impose constraints on the time evolution. We demonstrate that despite the time symmetric behavior of the microscopic constituents of the system, in many typical cases the external conditions lead naturally to the unidirectional macroscopic time evolution, independent of the internal interactions and how far the system is from equilibrium. We also illustrate our findings by direct calculation of the center of mass dynamics for a nonequilibrium steady state system partially free from external influence and for a system following the route to equilibrium.

\end{abstract}

\maketitle

%%%%%%%%%%%%%%%%%%%%%%%%%%%%%%%%%%%%%%%%%%%%%%%%%%%%%%%%%%%%%%%%%%%%%%%%%%%%%%%%%%%%%%%%%%%%%%%%%%%%%%%%%%%%%%%%%%%%%%%%

\section{Introduction}
\renewcommand{\theequation}{\arabic{equation}}

According to the microscopic laws of classical and quantum mechanics, all processes are expected to evolve similarly with time in one direction as they do in the opposite one. Such equivalence in the physical laws is usually called time reversal symmetry. However, macroscopic processes occur paradoxically in a preferred direction, i.e., the one with increasing entropy, as required by the second law of thermodynamics. This paradox, first introduced by Loschmidt, has been a matter of intense debate since the early days of statistical mechanics \cite{losch,edd,price,penrose,zeh}.    
The first attempt to explain the origin of this {\it arrow of time} was made by Boltzmann who established the idea of correlating the number microstates accessible to a system with its volume of phase space, and the equal {\it a priori} probability postulate: ``by removing some constraint from an isolated macroscopic system, the phase space volume available to it becomes fantastically enlarged, in such a way that it is extremely unlikely that the system will return to its initial state'' \cite{cerc,reif}. As a matter of fact, despite the groundbreaking consequences, this explanation demanded the introduction of new postulates, which are not directly connected to the fundamental laws of physics and, therefore, cannot be understood as a solution to Loschmidt's paradox. Yet, the notion of probability in terms of phase space volume is not applicable to systems far from equilibrium \cite{lebon}.     
  
After Boltzmann, many approaches were proposed to solve this conundrum, however, the majority of them relied on the {\it ad hoc} assumption of low entropy initial state \cite{penrose2,wald,cohen}. It has also been argued that the laws of physics must be modified to account for irreversibility \cite{ghi}, that entropy increasing and decreasing transformations may occur, but only those of the first case can be experimentally recorded \cite{maccone}. Other works suggest that the irreversible character of thermodynamic systems has a quantum mechanical origin \cite{pintos,rigol,linden,iyoda}, that it is related to the symmetry group of the Hamiltonian of the individual constituents \cite{bordel}, and that the thermodynamic and cosmological arrow of time are somehow connected \cite{layzer,gold}. Nevertheless, none of these results can be considered as a fully satisfactory resolution to the intriguing paradox. For a general review of the arrow of time problem, see Ref. \cite{halli}.

In this work, we study the Loschmidt's paradox by using a Newtonian approach, and the reasonable hypothesis that the pressure exerted on a fluid by the container's walls is a result of the reaction of the walls to the collisional forces applied by the constituents of the fluid. It is shown that, despite the time reversibility of Newton's laws of motion, when we apply such approach to observe the dynamics of the center of mass (CM) of a many-body system, irreversible transformations take place due to constraints imposed by interactions between the system and the surroundings. We also use this perspective to describe the dynamics of a diffusing gas in a steady regime and another reaching equilibrium. The results derived here have no restrictions with respect to the type or interactions between the constituents of the system, and how far the system is from equilibrium. The broad generality of the results presented here suggests that they may represent an alternative form of addressing the dynamics of a great variety of nonequilibrium thermodynamic systems.

\section{Newton's laws for single- and many-body systems}

Our aim in this section is to briefly review basic concepts of Newton's laws of motion to be used later in this work. Let us begin by summarizing each of the three laws. Newton's first law says that a body stays at rest or keeps moving at a constant velocity, unless a force acts upon it. To support the validity of this law, the existence of an inertial reference frame is assumed. Newton's second law states that the rate of change of the momentum of a body is directly proportional to the net force acting on it, and this change in momentum occurs in the direction of the net force. Finally, Newton's third law states that when two bodies interact by exerting forces on each other, the force exerted by the first body on the second is equal in magnitude and opposite in direction to the force that the second body exerts on the first.

If we consider Newton's second law for a single particle, it can be mathematically expressed in the well-known form \cite{feynman} 

\begin{equation}
\label{1}
\mathbf{F} = m \frac{d^{2} \mathbf{r} }{d t^{2}},
\end{equation}
where $\mathbf{F}$ is the force exerted on the particle, $m$ is its mass and $\mathbf{r}$ is the position vector. As can be seen, this equation is invariant for time reversal, i.e., under the substitution $t \rightarrow -t$, we can see that Eq.~\eqref{1} is unchanged due to the second time derivative. Mathematically, it signifies that if $\mathbf{r}(t)$ is a solution of Eq.~\eqref{1}, then so is $\mathbf{r}(-t)$, which means that, given any possible motion under Newton's laws, there exists another one in which the velocities are reversed.   

Let us now see what Newton's laws say for a system of many particles. First, we have that the localization of the CM of a system of $N$ particles is given by 
\begin{equation}
\label{2}
\mathbf{R}=\frac{1}{M}\sum_{i=1}^{N} m_{i} \mathbf{r}_{i}, 
\end{equation}
where $m_{i}$ and $\mathbf{r}_{i}$ are the masses and positions of the system constituents, respectively, and $M$ is the total mass. From the laws of motion stated above, it can be shown that the net external force exerted on the system is given as follows (Appendix A):
\begin{equation}
\label{3}
\mathbf{F}_{ext} = M\mathbf{a}_{cm},
\end{equation}
where $\mathbf{a}_{cm}$ is the  acceleration of the CM. This last equation is valid only if the total mass of the system is constant. It is difficult to overestimate the importance and generality of Eq.~\eqref{3}. But, we would like to remark three aspects of this result. First, we can conclude that a system of particles, whatever its physical state, subjected to no net external force, has the CM at rest or moving with constant velocity. The second point is that interactions occurring between constituents of the system do not interfere in the CM dynamics. The third point, which is fundamental for the discussion concerning nonequlibrium thermodynamic systems that we shall develop herein, is the fact that if we have some information about the net external force acting upon the system, by solving Eq.~\eqref{3} we can obtain the equation of motion of the CM. It is worth recalling at this early stage that the CM of a system of particles only represents an artificially created geometric entity without physical existence. However, it gives us a useful and simplified idea about how the mass of the system is distributed.

Although we have considered so far only the case of systems with total mass constant, it also valuable to mention what Newton's second law has to tell us about a  variable-mass system, namely, the situations in which the constituent particles can be exchanged with the external world. In such case, Eq.~\eqref{3} is no longer valid, and the applicable result is \cite{goldstein}
\begin{equation}
\label{4}
\mathbf{F}_{ext}=M\mathbf{a}_{cm}-\mathbf{u}\frac{dM}{dt},
\end{equation}
with $\mathbf{u}$ being the relative velocity of the escaping or incoming mass with respect to the CM of the system. From now on, we aim to use Newton's laws of motion to describe the dynamics of the CM of a system with a large number of particles in order to investigate the problem of irreversibility in macroscopic transformations.

\section{Center of mass dynamics and pressure field}

As mentioned in the previous section, Newton's laws of motion are time symmetric, which means that phenomena ruled by these laws are reversible. However, it is intriguing that the macroscopic behavior of nature has a preferred direction of time; a fact resumed by the second law of thermodynamics. In this section we start trying to conciliate these two conflicting ideas by using Newton's law, however, looking exclusively at the dynamics of the CM of a variety of thermodynamic systems. In other words, our objective here is to address Loschmidt's paradox under a Newtonian perspective of a system of particles. 

Our starting point is to derive an equation relating the resultant external force exerted on a fluid (system of particles) by the walls of the container with the pressure field $p$. Since we are not specifying the fluid dynamics and the shape of the container, the pressure field is in general a function of the spatial coordinates $x$, $y$ and $z$ and time $t$, i.e., $p=p(x,y,z,t)$. In Fig. 1, we see the general case in which we have a fluid contained in a recipient with arbitrary shape and a general pressure field. It is convenient to define the vector $d \mathbf{A}$ whose magnitude represents the area of the marked element of the surface of the recipient, whose direction is defined to be perpendicular to the surface element, according to Fig. 1. The vector $d \mathbf{F}_{ext}$ is the external force applied on the system caused by the respective surface element of the container's wall. Note that $d \mathbf{F}_{ext} (x,y,z,t)= - p(x,y,z,t)d \mathbf{A}$, where the minus sign appears due to the fact that $d \mathbf{F}_{ext}$ and $d \mathbf{A}$ are antiparallel, and $p(x,y,z,t)$ is a positive-definite field (See Sec. 3.4 of Ref. \cite{goldstein}). From now on we will omit the spatial and temporal dependence of the external force and pressure. Now, we can define the net external force exerted on the system as           

\begin{equation}
\label{5}
\mathbf{F}_{ext} = \int d \mathbf{F}_{ext} = - \int p d \mathbf{A},
\end{equation}
where the last integral is a surface integral, which means that it must be evaluated over the surface of the walls of the container. From Eqs.~\eqref{4} and ~\eqref{5} we can construct a relation between the acceleration of the CM of the system of particles and the pressure field on the surface of the container,
\begin{equation}
\label{6}
M\mathbf{a}_{cm}-\mathbf{u}\frac{dM}{dt} = - \int p d \mathbf{A}.
\end{equation}  

If we consider a system with fixed number of particles, $dM/dt=0$, we have that
\begin{equation}
\label{7}
\mathbf{a}_{cm} = - \frac{1}{M}\int p d \mathbf{A}.
\end{equation}
This equation tells us that the dynamics of the CM of the system of particles depends only on the pressure field on the surface of the container. Note that this result is valid independent of the nature of the interactions between the particles of the system, as long as the number of collisions per unit time between the particles and the walls is big enough for us to define a pressure field. About this last point, in considering a continuous pressure field as in Eq.~(\ref{7}), whose origin lies in the many individual constraint forces that takes place whenever a gas molecule collides with the wall, we are replacing a discrete sum of all constraint force contributions by an integral over the surface of the container (See Sec. 3.4 of Ref. \cite{goldstein}). This is quite reasonable in the case of a compressed gas with many constituent particles, given that there is a huge amount of collisions per unit time distributed over the wall surface. In this regime, fluctuations are negligible and the CM dynamics is not significantly influenced by individual collisions.  
\begin{figure}[htb]
\begin{center}
\includegraphics[height=1.7in]{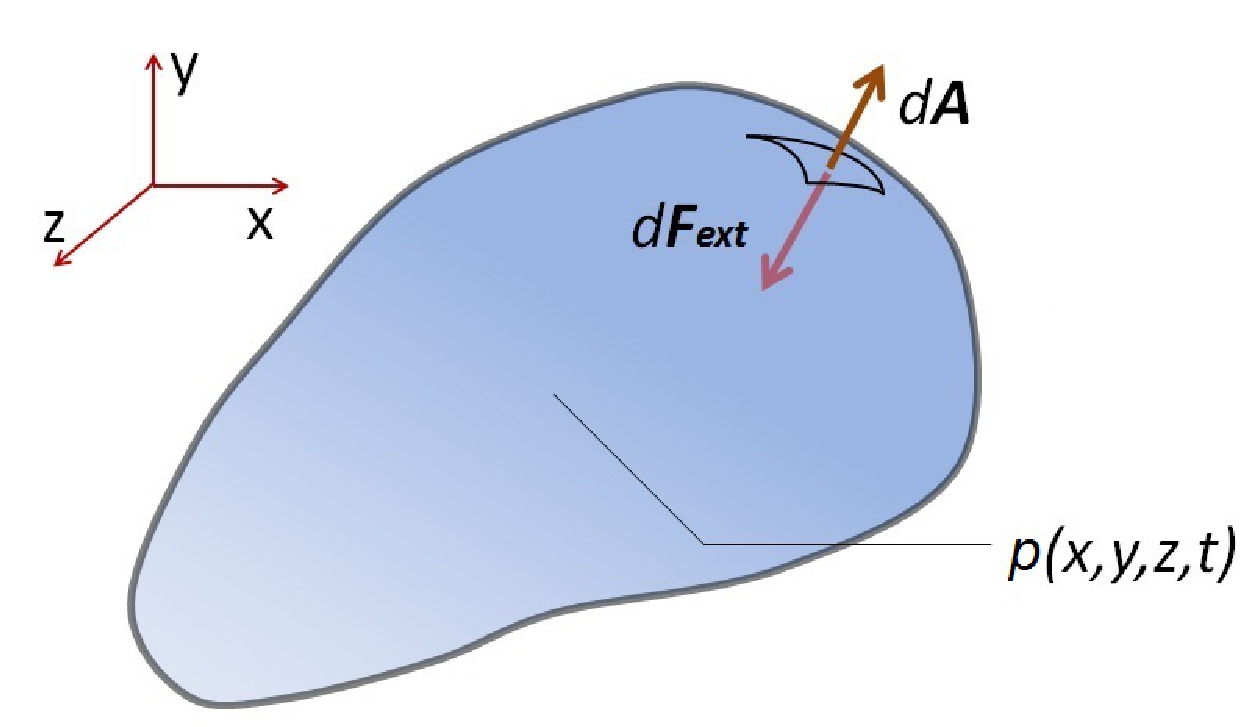}
\end{center}
\label{F1}
\caption{Contour surface involving a fluid with a pressure field $p(x,y,z)$. To each differential surface element we associate a normal outward vector $d \mathbf{A}$ where a differential external force $d \mathbf{F}_{ext}$ is exerted.}
\end{figure}

Before appreciating the importance of Eq.~\eqref{7}, let us first study the phenomenology behind it with the simplest non-trivial thermodynamic system: a gas at thermal and mechanical equilibrium with fixed number of molecules inside a box, as shown in Fig. 2a. Assuming that the molecules obey Newton's laws and that they are uniformly distributed in the available volume of the box, the pressure is obviously the same everywhere, including on the container's walls. The first point we have to keep in mind is that, since the gas is uniformly distributed in the box, and all molecules have the same mass, the CM of the gas is necessarily at the center of the box, as indicated in Fig. 2a. Second, if we consider the molecules of the gas as our system of particles, the external force applied on the system is uniquely due to the walls of the box when collisions between the molecules and the walls take place. However, it is easy to see from the symmetry of the box that the external force cancels out in all directions. Indeed, the left wall contributes with an external force pointing to the right, whereas the right wall contributes with an external net force of the same magnitude but in the opposite direction, rendering a zero external net force in the horizontal direction. The same analysis is applicable to the other opposite pairs of walls of the container. From this perspective, we can conclude from Eq.~\eqref{3}, with $\mathbf{F}_{ext} = 0$, that $\mathbf{a}_{cm} = 0$. That is to say, for a gas in equilibrium inside a box, the conditions imposed by the walls are such that the CM of the gas does not accelerate. Therefore, given that the CM is at rest due to the equilibrium conditions, it means that it will never move.

Now, let us apply the result found in Eq.~\eqref{7}. Since we have a gas in equilibrium, the pressure field is constant all over the surface of the container (neglecting gravity effects). Also, if the container is closed, we have to evaluate the integral over a closed surface. Thus,
\begin{equation}
\label{8}
\mathbf{a}_{cm} = - \frac{p}{M} \oint d \mathbf{A} = 0,
\end{equation}
which is accordance with our discussion above. However, it is important to mention that the result of Eq.~\eqref{8} holds, independently of the shape of the box, because $\oint d \mathbf{A} = 0$ for any closed surface. From now, we shall make use of Eqs.~\eqref{6} and~\eqref{7} to address the problem of irreversibility at the macroscopic scale.

Before exploring how Eq.~(\ref{7}) applies to many interesting cases of dynamical many-body systems to be addressed here, it is convenient to mention some important aspects related to the physical context in which the pressure field $p$ arises in many branches of physics. Historically, Newton himself used the concept of pressure (not as a scalar field as presented here) dozens of times in his monumental 1687 work, ``Philosophiae Naturalis Principia Mathematica'', to illustrate how the surfaces of a vessel are capable of sustaining a fluid inside it \cite{newton}. In fluid mechanics, the Navier-Stokes equation, which can be seen as Newton's second law of motion for fluids \cite{landau}, already presents the pressure as a continuous scalar field, in much the same way as we mathematically introduced here. In turn, in modern classical mechanics (see Sec. 3.4 of Ref. \cite{goldstein}), and in the kinetic theory of gases \cite{pauli}, both serving as a support for classical statistical mechanics \cite{reif}, the concept of pressure as a scalar field can also be employed, however, defined upon a molecular view of matter. In fact, all post-Newtonian theories assign the macroscopic view of the pressure exerted by a container's wall on an enclosed fluid to the ``reaction of the wall to the collision forces exerted by the atoms on the wall'' \cite{goldstein}, a fact which unavoidably requires a statistical treatment to be properly analyzed. Overall, our modern classical view of pressure is a result of both Newton's theory of motion and the idea that matter is made of atoms. As we know, the latter idea was established a couple of centuries after the former. At this stage, we close this section by remarking that: i) whatever classical theory one wants to use, Newtonian mechanics will necessarily play a central role in the definition of pressure, and that ii) this definition alone cannot explain the emergence of irreversible phenomena in nature. This second remark is essential for the forthcoming analysis.

\section{Connection between Newton's Laws and irreversible processes}   

Now we start addressing the arrow of time problem with basis on the formalism derived in the last section. The simple example of the gas in equilibrium of Fig. 2a leads to a number of restrictions to the possible evolution of the system. For example, based on what was shown, we can say that it is impossible that the gas by itself can evolve spontaneously to a state in which all molecules are confined in one half of the box, as in Fig. 2b. In fact, this is because a gas which occupies one half of the volume has the CM displaced with respect to the case of Fig. 2a. But, since the CM is at rest and $\mathbf{F}_{ext}=0$ in the first case, the second situation can never be attained. Conversely, it is possible that the gas in the second case can evolve to the first because the net external force is nonzero and points to the center of the container, where the CM of the equilibrium state is localized. It is important to emphasize that so far our arguments do not indicate that the second case will evolve to the first one, but that there are no restrictions for such evolution. However, an evolution from the first state to the second is strictly prohibited. The fact that $\mathbf{F}_{ext} \neq 0$ in the second case is because the external force in the horizontal direction is only due to the left wall of the container, or, more formally, the right wall does not contribute to the integral of Eq.~\eqref{5} ($p=0$). On the other hand, on the left wall, where $p > 0$, we have a net external force pointing to the right ($d \mathbf{A}$ points to the left).    
\begin{figure}[htb]
\begin{center}
\includegraphics[height=2.7in]{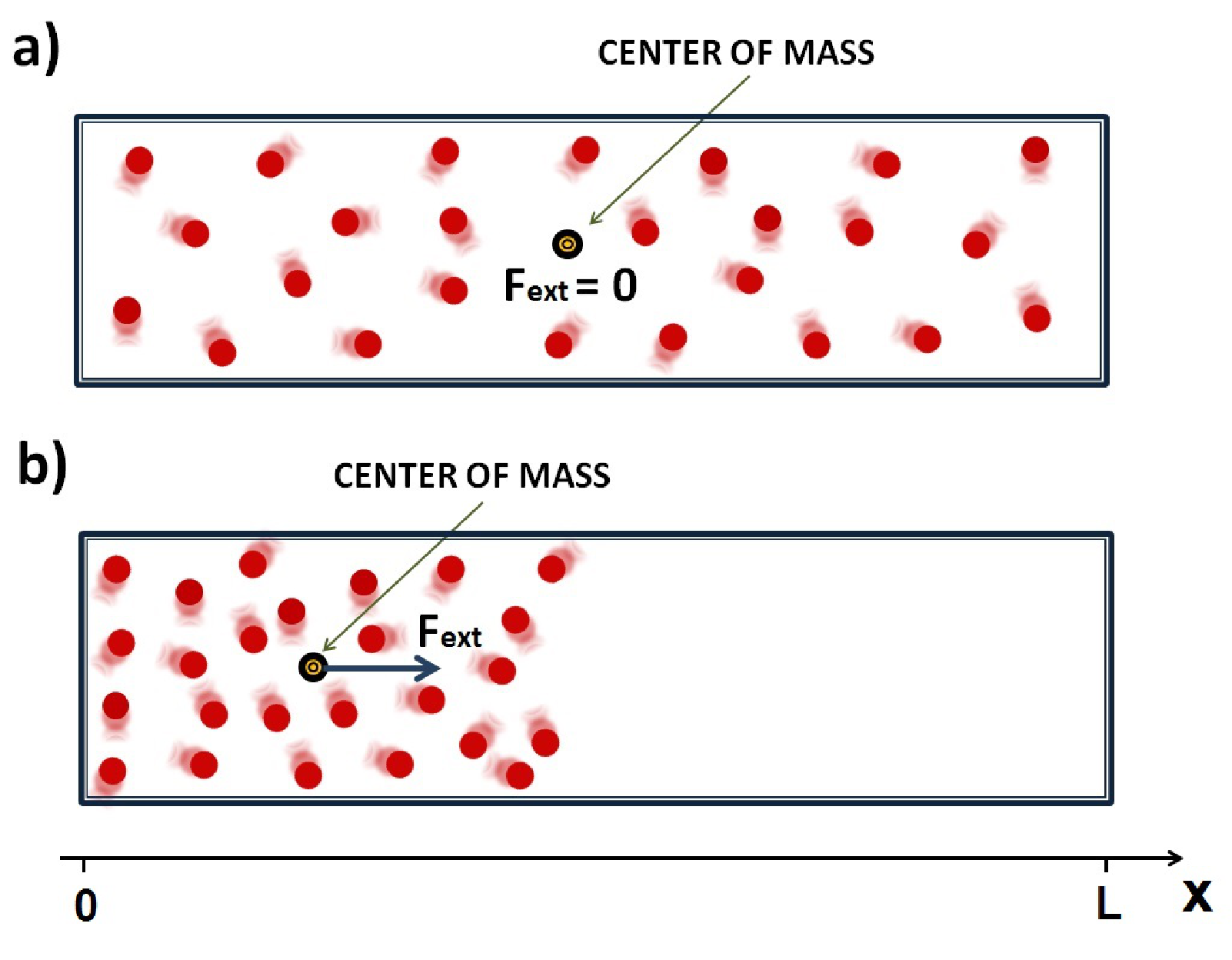}
\end{center}
\label{F1}
\caption{Gas of point particles inside a box. In a) we have the gas uniformly distributed over the volume, whereas in b) we have the situation in which the gas occupies only the left half of the box. The net external force due to the momentum transferred to the particles by the walls during collisions is shown at the position of the CM.}
\end{figure}

At this point, it is worthwhile to make a comparison between our conclusions for the problem presented above and the ones we obtain from classical statistical mechanics for a system of point particles, due to Boltzmann and Gibbs \cite{mackey}. In that approach, the state of a gas is described by its {\it macrostate}, which is specified by the macroscopic properties such as energy, volume, temperature, etc. To each macrostate, there are many distinct {\it microstates} associated, which in this case correspond to the arrangements of the position and momentum of the $N$ particles. That is to say, each microstate corresponds to a point in a $6N$-dimensional phase space for a system with a prescribed macrostate. In this view, the idea is that the number of microstates related to the configuration of Fig. 2a is enormously larger than the number of microstates associated with the configuration of Fig. 2b. Then, if we assume that all microstates accessible to the system are equiprobable over a long period of time, the {\it ergodic hypothesis}, we expect that the system can naturally evolve from the state of Fig. 2b to the state of Fig. 2a, however, the contrary transformation would have a very small, but nonzero, probability to occur. Conversely, if we consider the present approach, the latter transformation would never happen, since the net force is null in the case of Fig. 2a, and for any displacement of the CM from the equilibrium position, a restoring force tends to bring the CM back to the original point, as shown in Fig. 2b. This restoring force is due to the imbalance of the external forces on the left and right walls of the container. From now on, we shall apply the ideas developed so far to study the irreversibility of molecular gas dynamics in some key situations in order to give quantitative support to our arguments.
 
\subsection{Gas diffusion in a box}

It is valuable to see how our ideas are applied to the case of a gas diffusing inside a box. In this situation, shown in Fig. 3, a gas initially confined on the left side of a box is allowed to diffuse through the available volume. Let us first consider the simplest situation of an infinitely long box, $L \rightarrow \infty$. Since the external force along the horizontal direction is due uniquely to the momentum transferred by the left wall to the gas molecules (the external force cancels out along the other directions), the net external force must point to the right according to Fig. 3. It is important to note that, once there is no influence of the right wall in this configuration, it is impossible to have the net external force pointing to the left. As a conclusion, the dynamics of this system of particles is such that its CM can only move to the right, and the contrary never occurs. This constraint imposed by the dynamics of the CM prohibits, for example, the system of evolving from a configuration in which the molecules are spread out over the reservoir, with the CM far from the walls, to the initial state in which the molecules are arranged close to the left wall. However, the inverse evolution is perfectly permitted by this constraint. In this sense, the analysis of the dynamics of the CM, which is formulated with basis on Newtonian arguments, imposes an asymmetry for the time evolution of the system in accordance with the second law of thermodynamics.
\begin{figure}[htb]
\begin{center}
\includegraphics[height=1.3in]{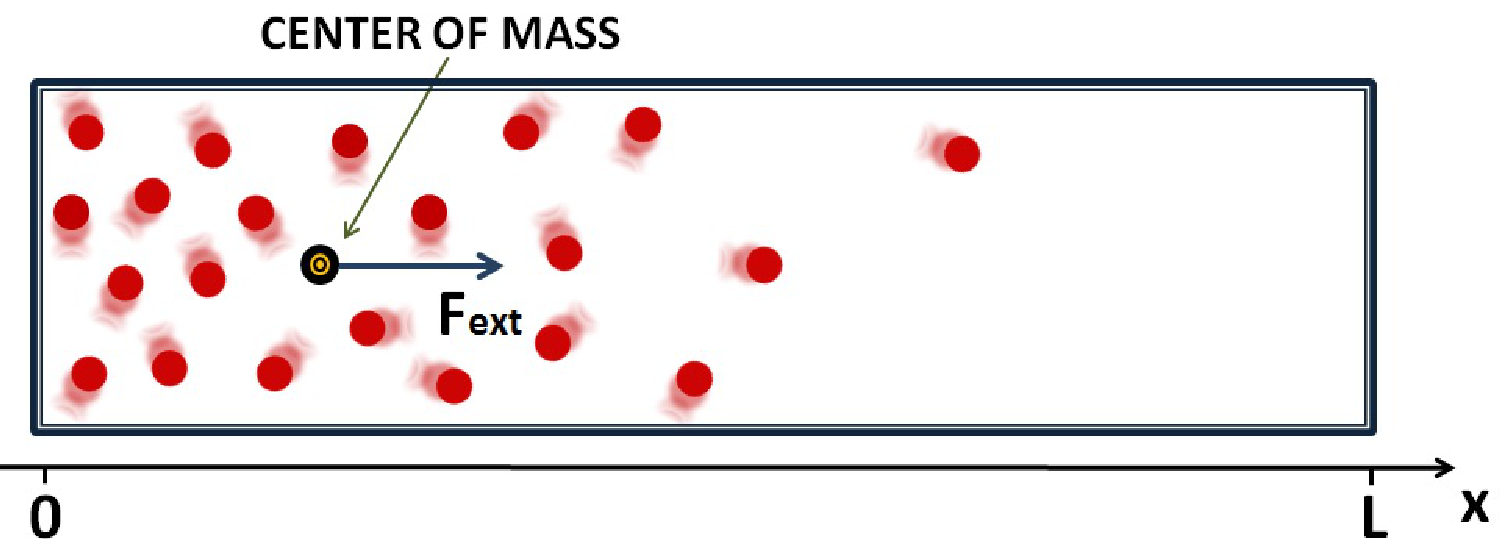}
\end{center}
\label{F1}
\caption{Molecules of a gas diffusing through a box of length $L$. The net external force due to the collisions of the molecules with the walls is shown at the position of the CM.}
\end{figure}

As mentioned before, the arguments concerning the CM dynamics are independent of the type of interaction between the system constituents. However, to give an insight into this dynamics, let us consider the system as an ideal gas. In this case, the net external force, which is proportional to the pressure on the left wall, can be assumed to be inversely proportional to the distance $x$ of the CM from the wall (at $x = 0$), for some cases of interest (Appendix B). Thus, we can write the magnitude of the external force as
\begin{equation}
\label{9}
F_{ext}(x)= \frac{f(T,N)}{x},
\end{equation} 
where $f$ is a function of the temperature $T$ and the number of particles $N$. In this form, by using Eq.~\eqref{3}, the dynamics of the CM is given by the second-order nonlinear ordinary differential equation $\ddot{x}=f/Mx$. By making $f=1$ (temperature at the left wall and number of particles fixed) and $M=1$ for simplicity, we sketch the behavior of the position of the CM as a function of time for $x(0)=x_{0}$ and $\dot{x}(0)=0$, Fig. 4.        
\begin{figure}[htb]
\begin{center}
\includegraphics[height=1.26in]{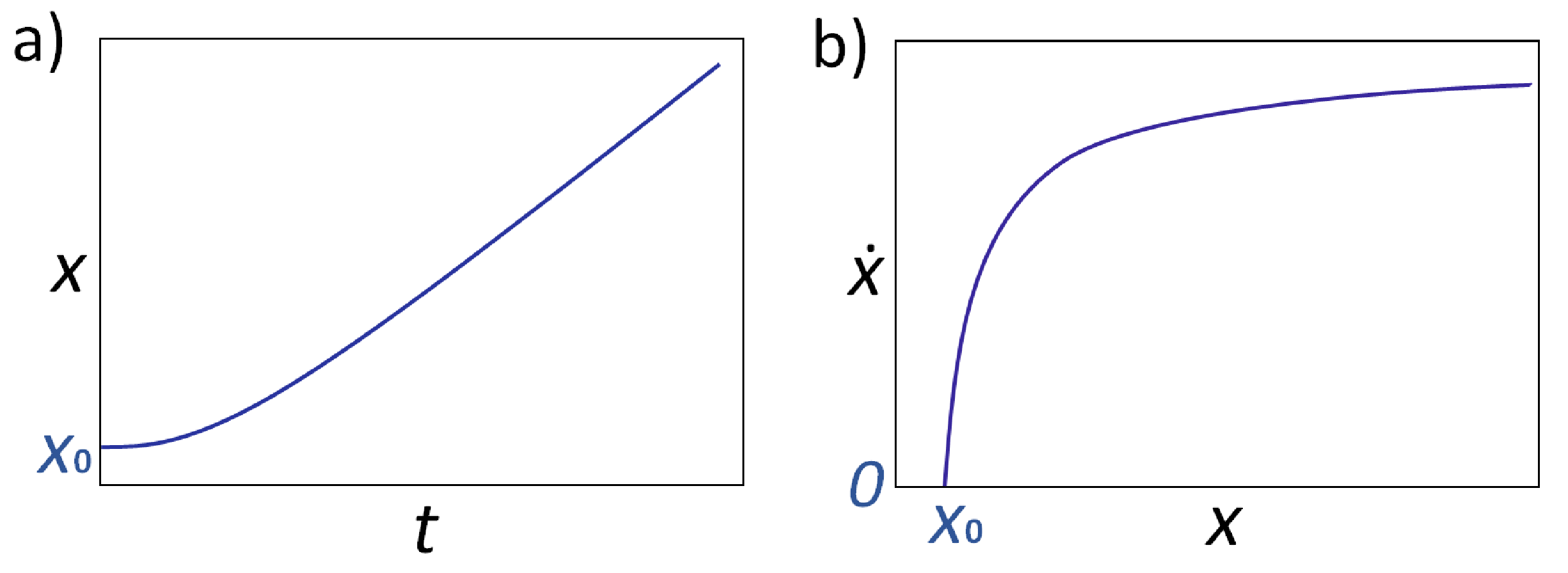}
\end{center}
\label{F1}
\caption{Dynamics of the CM of a diffusing ideal gas in an infinitely long box initially at $x_{0}$ and at rest, $\dot{x}(0)=0$. a) Graph of the position as a function of time. b) Graph of the velocity as a function of the position.}
\end{figure}

As can be seen in Fig. 4a, the CM of the diffusing gas is always moving away from the left wall. It has an initial increase in velocity which stabilizes after the CM is sufficiently far from the left wall. Physically, the initial increase in velocity is due to the nonzero acceleration acquired by the CM due to the external force applied by the box because of the collisions between the molecules and the left wall. This force diminishes with time because the rate of collisions with the left wall decreases as the CM moves to the right. The velocity acquired by the CM becomes constant because the collisions with the left wall cease after some time. This steady state behavior could only be modified when some molecule reaches the right wall, which is a case that we are not considering now. The graph of the velocity as a function of the position is also in agreement with our argumentation, Fig. 4b.

Now we want to consider the case of a box with finite length. In this case, the influence of the right wall in Fig. 3 has to be considered. Similar to the previous case, each wall exerts on the gas a repulsive force inversely proportional to the distance to CM. This is the net effect for the case in which the CM of the gas is at rest inside the box. However, if the CM is moving against a given wall, this latter has to impose an extra momentum to invert the motion of the molecules along the perpendicular direction. This fact renders a term proportional to the velocity of the CM (see Appendix C). Putting all this together, we have that the external force exerted by both the left and right walls along the $x$ direction yields    
\begin{equation}
\label{10}
F_{ext}(x,\dot{x})= \frac{f(T,N)}{x} + \frac{f(T,N)}{x-L} - g(T,N) \dot{x},
\end{equation} 
where the damping factor $g$ is a function that depends both on the temperature and number of particles. Accordingly, the dynamics of the CM is obtained by the following second-order nonlinear ordinary differential equation
\begin{equation}
\label{11}
\ddot{x} = \frac{1}{x} + \frac{1}{x-L} - \dot{x},
\end{equation}
where we have made $M=1$, $f=1$ (in this case the temperature at the left and right walls must be kept constant, possibly by using an external thermal bath) and $g=1$ for simplicity. The behavior of the solution of this equation for $x(0)=x_{0}>L/2$ and $\dot{x}(0)=v_{0}>0$ is shown in Fig. 5a.
\begin{figure}[htb]
\begin{center}
\includegraphics[height=1.26in]{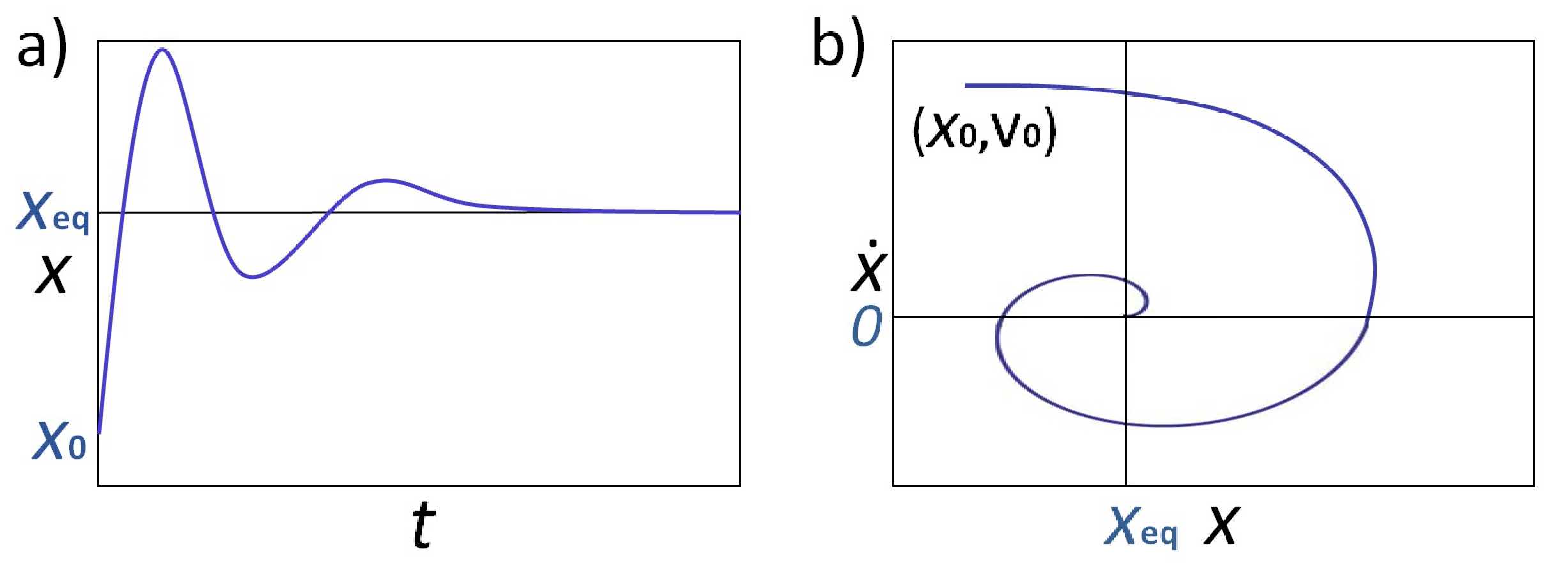}
\end{center}
\label{F1}
\caption{Dynamics of the CM of an ideal gas in a finite box. The initial and final positions are $x_{0}$ and $x_{eq}$, respectively. The initial velocity is $v_{0}$ and the final is $0$, as expected. a) Graph of the position as a function of time. b) Graph of the velocity as a function of the position. Both curves evidence the route to equilibrium.}
\end{figure}

The solution shows that the CM of the gas oscillates with the amplitude gradually decreasing to an equilibrium position $x_{eq}=L/2$, i.e., after a transient time the CM reaches its equilibrium position at the center of the box, which is the expected result if one considers thermodynamic arguments. However, we emphasize again that we used only a Newtonian scenario to derive this result, and the assumption that the pressure exerted by the walls of a reservoir on a fluid is a result of the reaction of the walls to the collisional forces applied by the particles of the fluid. The behavior of the velocity of the CM as a function of the position is shown in Fig 5b. The gradually decreasing oscillation can also be observed in this analysis. In both cases studied in this subsection, we analyzed only the dynamics of the CM in the $x$ direction. However, it is easy to see that the extension of this method to cases in two and three dimensions is straightforward. For this reason, in all examples in this article the dynamics is analyzed only in one dimension, without loss of generality.   

It is worth mentioning that such behavior of the CM in pursuing a steady motion or an equilibrium point only took place because of the large number of constituents involved. In fact, in the limit case of a single particle inside a box, the CM motion would be that of the particle, i.e., bouncing around in the available volume.     
The crucial difference between the dynamics of a single particle (or few particles), and the CM dynamics of a many-body classical system is that in the first case the body will experience forces isolated in time, which are few and far between. The forces involved in such time distribution are not capable of carrying the system to an equilibrium position, in the absence of dissipation, because they act one at a time, without the possibility of cancelation. On the other hand, in the second case the CM is ``experiencing'' simultaneously several forces pointing in infinitely many directions. This configuration of forces, as we have shown, provides the possibility of a resulting force pointing in a single direction and, eventually, cancelation. This is the reason why the CM is only allowed to attain an equilibrium situation when the system has infinitely many constituents. Also in this context, we should not be concerned about the fact that the CM reaches an equilibrium position in the absence of dissipative forces acting on the constituents. This is because the CM is not a physical system, but only an imaginary geometric point in space.

\subsection{Diffusion of a gas from the center of an infinitely long box}

As the name suggests, the diffusion in an infinite region has to do with the case of a gas initially concentrated in a small region in space, say around $x=0$, which is later allowed to diffuse along the negative and positive directions of the $x$-axis. To illustrate this case, let us use the box in Fig. 6 and assume that it is infinitely long, $L \rightarrow \infty$. In this regime, the left and right walls have no effect in the sense that there are no collisions between the molecules and these walls. In this regard, no net external force takes place in the horizontal direction, so that we are led to conclude from Eq.~\eqref{3} that the CM, which is initially at rest, will remain so forever, independent of how the molecules move. This information tell us, for example, that it is impossible that the molecules of the gas can move mostly to either the positive or negative direction, once the CM must necessarily stay at $x=0$, Fig. 6a. However, this fact cannot give us information about how the density of molecules evolves with time. In this context, extra information can be obtained if we mentally divide the gas into two parts: the molecules contained on the left side of the box (subsystem $A$), and the molecules contained on the right side (subsystem $B$). The subsystems $A$ and $B$ were separated in Fig. 6b by the imaginary vertical dashed line.
\begin{figure}[htb]
\begin{center}
\includegraphics[height=2.6in]{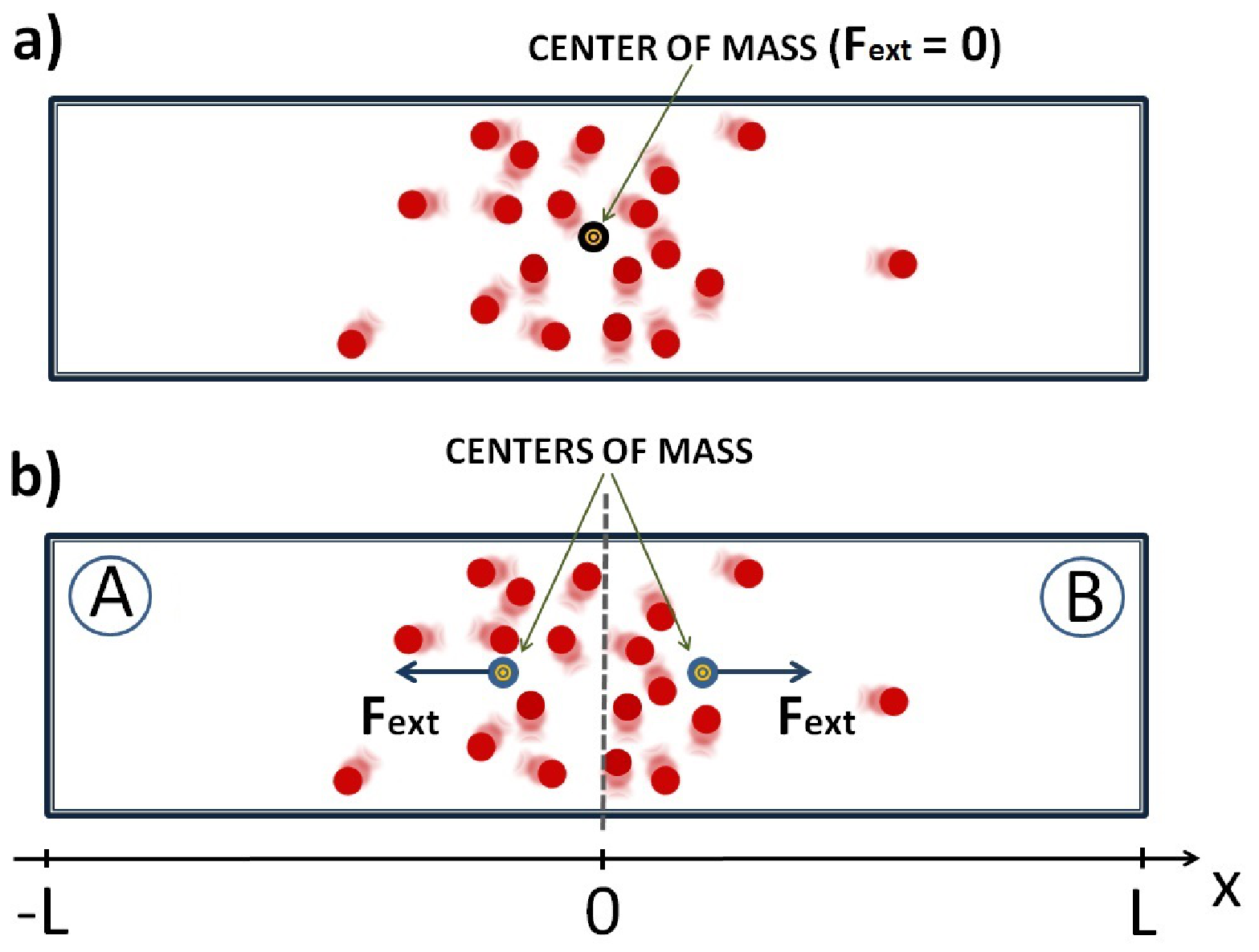}
\end{center}
\label{F1}
\caption{Diffusion of a gas starting from the center of a long box. a) The CM of the whole system does not move due to the absence of a net external force. b) If we mentally divide the system into two subsystems A and B, each occupying one half of the reservoir, information about the dynamics of the molecules on each side can be obtained if we analyze the centers of mass of the subsystems.}
\end{figure}

Let us analyze, for example, the dynamics of the CM of subsystem $B$. First, note that the number of particles of each subsystem varies with time because particles can be exchanged between them. Then, Eq.~\eqref{4} must be used to study the dynamics of the CM of subsystem B. Nevertheless, if we consider that the initial molecular distribution is symmetric around the imaginary dividing line, we can consider $dM/dt\approx 0$ in Eq.~\eqref{4}, so that Eq.~\eqref{3} can be used in this case. Second, contrary to the whole system, subsystem $B$ experiences a nonzero net external force in the horizontal direction because of the momentum transferred by the molecules of subsystem $A$ to the molecules of subsystem $B$ due to the collisions that occur at the imaginary dividing line. For symmetry reasons, this net external force must point to the right as indicated in Fig. 6a, which means that the CM of subsystem $B$ must move to the right. By the same token, the CM of subsystem $A$ has to move to the left in such a way that the CM of the whole system stays at rest. This last analysis provides new information that could not be obtained by looking at the system as a whole. In fact, from the analysis of the subsystems we can conclude that the complete system cannot stay confined in the region around $x=0$, since the centers of mass of subsystems $A$ and $B$ must move away from each other as shown in Fig. 6b. Also, the movement is accelerated as long as there exist collisions between the subsystems. After the collisions cease, their CM will move individually in opposite directions at constant velocity. In this last case, the dynamics of the individual CM becomes similar to that shown in Fig. 4. In a similar fashion, if the box of Fig. 6 is not infinitely long, the effect of the left and right walls cannot be neglected, and the dynamics of the CM of subsystems $A$ and $B$ must be similar to that shown in Fig. 5, that is, the centers of mass will oscillate symmetrically with gradually decreasing amplitude around an equilibrium position; presumably at $x=-L/2$ and $x=L/2$, respectively, as expected from the second law of thermodynamics.

\subsection{Interaction between hot and cold gases}

The present ideas can also be applied to systems with non-uniform temperature. Here, we shall apply them to the case of two interacting gases at different temperatures. Such system, commonly used to study the Maxwell's demon paradox \cite{max,goold}, consists in a box with a partition separating a hot and a cold gas, which, after removing the partition, are allowed to exchange particles with each other, as shown in Fig. 7. The hot gas is initially confined on the left side of the box and the cold gas on the right side. If we assume that both gases have the same type of molecules uniformly distributed in space, the CM is localized at the center of box, as indicated.
\begin{figure}[htb]
\begin{center}
\includegraphics[height=1.22in]{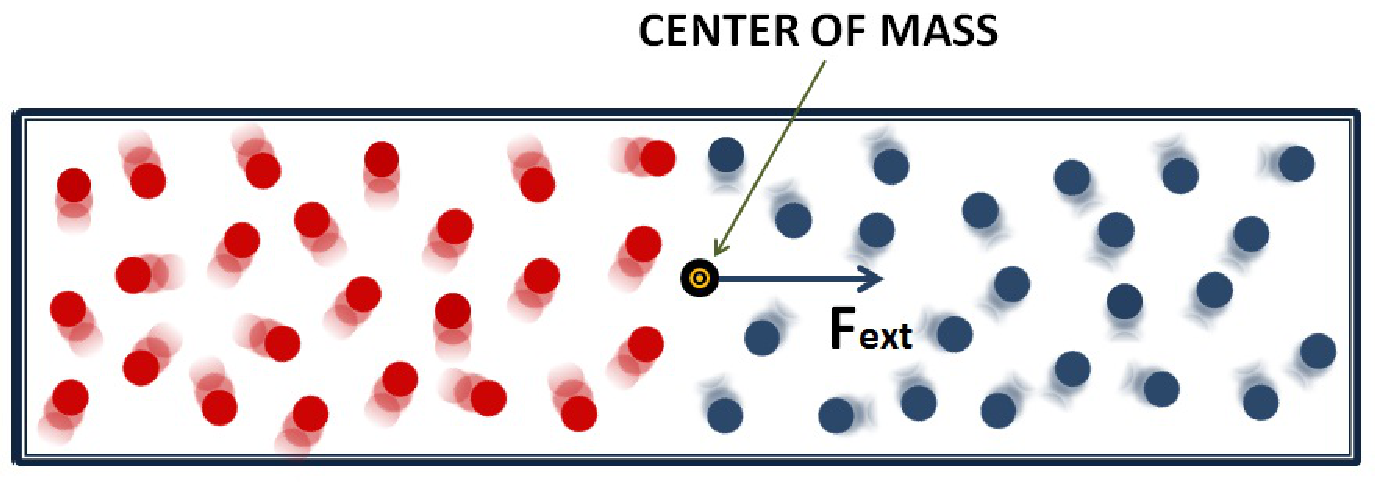}
\end{center}
\label{F1}
\caption{Interaction between hot (left) and cold (right) gases uniformly distributed in a finite box. Because of the higher temperature on the left side, there is a higher transfer of momentum from the left wall to the system, when compared to the right wall. This entails a net external force pointing to the right, as shown at the position of the CM.}
\end{figure}

In order to study the evolution of this system by using our arguments, we must concentrate on the transfer of momentum by the recipient to the gas at the left and right walls due to collisions. If the gas on the left side is hotter than the gas on the right, the average momentum transferred from the left wall to the molecules is greater that of the right wall. Therefore, the net external force on the gas must point to the right, as shown in Fig. 7. As a consequence, the CM, which is initially at rest, moves to the right due to the acquired acceleration. For the CM to move to the right, it is necessary that molecules of the left side, the hotter ones, move to the right side to join the colder ones. In this respect, the net result is that we have a heat flow to the right so that the gas on the right side becomes hotter, whereas the gas on the left becomes colder. This is in perfect agreement with the Clausius statement of the second law of thermodynamics \cite{clausius}: ``Heat can never pass from a colder to a warmer body without some other change, connected therewith, occurring at the same time''. Indeed, from the constraint that the CM of the system must move to the right, we cannot have a net flux of hot particles moving towards the left wall. Our conclusion can also be seen as a molecular manifestation of the so-called Soret effect, or thermal diffusion, in which there is a mass flow establishment as a consequence of a temperature gradient \cite{duhr,rah}. 

During the thermodiffusive stage, the tendency is to equalize the temperatures on both sides of the box, due to the net flux of hot particles to the right, which entails a higher concentration of particles on the right side. However, such imbalance in the concentrations will contribute with a net external force pointing to the left, according to the arguments presented in subsection IV. A. Therefore, the system will tend to uniformize the temperature and concentration of particles, as expected from the second law.

\section{Discussion and conclusions}

At this stage, we are in a position to argue how the fundamental ideas concerning Loschmidt's paradox can gain a renewed perspective if addressed from a Newtonian viewpoint for a system of many bodies according to the reasoning developed here. As already discussed, this paradox poses the following question: how can we obtain irreversible phenomena from reversible time-symmetric physical laws? In the case of Newton's laws, after the variety of examples shown in section IV, we conclude that, despite the laws are dynamically invariant from moving forward to backward in time, which establishes time-reversal symmetry to the molecular motion, if we use these same laws to make a macroscopic analysis of a many-body system paying attention to the CM, we see that there is a time-reversal symmetry breaking in the evolution of the whole system. In the present work we could understand that the symmetry breaking has its root in the external forces acting on the system. This influence depends, for example, on whether the system is contained in a closed or in an open container, if it is free from external interactions, etc. In all cases analyzed here, allowed and prohibited evolutions were found for the many-body systems by using a Newtonian reasoning, and the fact that the pressure field exerted on a fluid has its origin from the forces of reaction imparted by the container's walls due to the collisions with the particles of the fluid. Such results would not be accessible if we had analyzed the individual motion of the molecules, as these problems are normally approached. In general terms, irreversibility came into play only when we used Newton's laws to study the dynamics of the CM of the system. This is how the arrow of time shows up.

In this context, not less important is the question: consider a classical system whose state is represented by a point in phase space over a trajectory representing its evolution. If we invert the velocities of all particles in that point, and let the system evolve, would we obtain the former initial point of the trajectory with all velocities reversed? We have no reason to give a negative answer to this question, even knowing that if this were possible the entropy of the system would decrease, in apparent conflict with the second law. But how to conciliate this possibility with the prohibited evolutions found when we discussed the examples of the previous section? The answer lies in the fact that the act of reversing the velocities of the particles requires an external force whose direction is contrary to that imposed by the natural external conditions, like the ones shown in the previous section. Namely, the effect of the reversal in the molecular velocities is to decrease the entropy up to a minimum value of the trajectory in phase space \cite{me}. After this moment, the entropy must increase according to the external influence exerted on the system, as discussed before.   

Another important point to call attention is that the well behaved dynamics of the CM found in all examples above is because we implicitly considered the systems as formed by many particles so that the rate of collisions between the container's walls and the particles is high enough for us to consider that the net external force on the system is approximately constant for molecular time scales. If this condition is relaxed, for example, in systems of rarefied gases in which the dimensions of the container are of the order of the mean free path of the particles, the net external force on the system is significantly perturbed whenever a collision between a constituent particle and some of the walls occurs. In this latter case, the motion of the CM of the system is erratic and, as a consequence, the unidirectional evolution shown in our examples is no longer sustained. Another consequence is that equilibrium conditions could not be attained in this case. Under this viewpoint, we can see why irreversibility takes place exclusively in many-body systems.     

To gain further intuition on the reason why the CM of the many-body systems analyzed here attains an equilibrium configuration, specifically in the case of the diffusion in a box, it is insightful to recall the results obtained from the theory of dynamical billiards, developed mainly by Bunimovich and Sinai \cite{sinai}. A dynamical billiard consists in a particle that moves alternating between straight line trajectories with constant speed and specular (elastic) reflection from the boundaries of a closed space region, whose potential is designed to be zero, without any dissipation of energy. It has been demonstrated that for many boundary shapes, the possible trajectories of few confined particles are ergodic, meaning that after a long period of time the particles independently visit all the available space uniformly. As a consequence, for a system with a large number of particles as a gas, the tendency is to have particle-wall collisions distributed uniformly along the boundaries. This effect renders a cancellation of the external forces from the boundaries, hence leading the CM of the system either to a stationary (constant speed) state, as in the case of the infinite box (see Fig. 4), or to rest as in the case of the finite box, Fig. 5. This is valid although the perpetual stochastic motion of each individual constituent particle.  
 
In conclusion, we introduced a possible form of reconciling the time-reversal symmetry of Newton's laws with the macroscopic irreversibility observed in thermodynamic processes with basis on the dynamics of the CM of a many-body classical system. In particular, we showed that in some cases the analysis of the CM dynamics can provide valuable information related to the evolution of the system, as indicated in some selected examples. In such cases, it was found that the conditions imposed by the external influence on the system, which naturally impacts on the CM motion, are decisive for the unidirectional evolution of the system. In this regard, we consider that these findings provide important insights into the Loschmidt's paradox from a classical perspective and shed a new light on the study of nonequilibrium phenomena. Of course we cannot draw general conclusions from only these few examples and say that the paradox is solved. However, we believe that a promising strategy to be adopted is to carry out a similar analysis for more complex systems, for example, with internal degrees of freedom and boundary conditions different from the ones presented here. In future works we intend to investigate the possibility of such extensions.

\section*{ACKNOWLEDGEMENTS}
The author acknowledges financial support from the Brazilian funding agencies Coordenação de Aperfeiçoamento de Pessoal de Nível Superior (CAPES, Finance Code 001), and Conselho Nacional de Desenvolvimento Cientfico e Tecnológico (CNPq, Grant No. 303451/2019-0).

\section*{APPENDIX A: DYNAMICS OF THE CENTER OF MASS OF A SYSTEM OF FIXED MASS}
\renewcommand{\theequation}{A-\arabic{equation}}
\setcounter{equation}{0}

Here, we shall describe the dynamics of the CM of a system of particles of fixed total mass $M$. First, we have that the velocity of the CM is given by the time derivative of the position vector given in Eq.~\eqref{2},
\begin{equation}
\label{12}
\mathbf{V} = \frac{d \mathbf{R}}{dt} = \frac{1}{M} \sum_{i=1}^{N} m_{i} \mathbf{v}_{i},
\end{equation}
where $\mathbf{v}_{i}$ is the velocity of the $i$th particle of the system. This equation can be rewritten in terms of the momenta of the particles $m_{i}\mathbf{v}_{i}$ as
\begin{equation}
\label{13}
M\mathbf{V} =  \sum_{i=1}^{N} m_{i} \mathbf{v}_{i}.
\end{equation}
Therefore, we have that the total momentum of the system is equal to the product between the total mass and the velocity of the CM. 

Now, by using Newton's second law, if we differentiate Eq.~\eqref{13} with respect to time, we obtain an expression relating the total force acting on the system $\mathbf{F}_{tot}$ and the acceleration of the CM $\mathbf{a}_{cm}$, 

\begin{equation}
\label{14}
\mathbf{F}_{tot} = \sum_{i=1}^{N} \mathbf{F}_{i} =  \frac{d (M \mathbf{V})}{dt} = M \frac{d \mathbf{V}}{dt} = M \mathbf{a}_{cm},
\end{equation}
where $\mathbf{F}_{i}$ is the resultant force on the $i$th particle. The resultant forces may include both forces internal to the system, due to the interaction with other constituents, and external forces, due to elements outside the system. Nevertheless, Newton's third law says that the internal forces taking place between any pair of particles of the system are equal in magnitude, but opposite in direction. Thus, when the sum over all internal forces in
Eq.~\eqref{14} is realized, they must cancel in pairs so that the net force on the system is due uniquely to external forces. In this form, Eq.~\eqref{14} yields
\begin{equation}
\label{15}
\mathbf{F}_{ext} = M\mathbf{a}_{cm},
\end{equation}
which is the result of Eq.~\eqref{2}. Physically, this equation says that the CM of a system of particles moves exactly as a particle of mass $M$ would move if it were under the influence of the net external force acting on the system.

\section*{APPENDIX B: EXTERNAL FORCE EXERTED BY A WALL ON AN IDEAL GAS WITH THE CENTER OF MASS AT REST}
\renewcommand{\theequation}{B-\arabic{equation}}
\setcounter{equation}{0} 

Now, we want to show the dependence of the external force exerted by a wall with the inverse of the distance between the wall and the CM of an ideal gas system, used in Section IV. In doing so, we will study two cases. The first one is when the gas is uniformly distributed around the container's wall and the other when the gas has a Gaussian distribution of particles along the direction perpendicular to the wall. This last case is better suited in diffusive processes. For now, we assume that the CM is at rest. We relax this condition in Appendix C.    

For the first case, let us consider the case of a gas that occupies uniformly part of a container, which had the partition quickly removed as shown in Fig. 2b. In this case, the magnitude of the external force exerted by the left wall is given by $F_{ext} = p A$, where $p$ is the uniform pressure of the gas and $A$ is the area of the wall. By using the ideal gas law, $pV = Nk_{B}T$, with $N$, $k_{B}$ and $T$ being the number of particles, the Boltzmann constant and the absolute temperature, respectively, and $V$ the volume occupied by the gas, the external force becomes  
\begin{equation}
\label{16}
F^{(0)}_{ext} = \frac{Nk_{B}T}{V}A = \frac{Nk_{B}T}{\alpha L}= \lambda k_{B}T ,
\end{equation}
where we assumed for simplicity that the space occupied by the gas is a rectangular parallelepiped of volume $V=\alpha AL$, where $L$ is the horizontal length of the container and $\alpha$ is the fraction of the total volume of the container occupied by the gas. The parameter $\lambda = N/ \alpha L$ is the linear density of particles in the region occupied by the gas along the $x$-axis. We used the superscript $0$ to indicate that this external force accounts for the case in which the CM is at rest. Since the gas is uniformly distributed in the occupied volume, we have that the position of the CM in the $x$-axis is $x_{cm}= \alpha L/2$. Thus, we have that
\begin{equation}
\label{17}
F^{(0)}_{ext} = \frac{Nk_{B}T}{2x_{cm}}.
\end{equation}  
This result shows the dependence of the external force exerted by the left wall with the inverse of the distance to the CM.

We now consider the case in which the distribution of $N$ particles along the positive $x$ direction is Gaussian, with the CM instantaneously at rest, with the highest density at $x=0$, which coincides with the position of the left wall. Fig. 3 is a good illustration of this case. The distribution is given by
\begin{equation}
\label{18}
\lambda(x) = \sqrt{\frac{2}{\pi}} \frac{N}{\sigma} \exp \left({\frac{-x^{2}}{2 \sigma^{2}}} \right), \qquad (x>0)
\end{equation} 
where $\sigma$ quantifies the broadness of the distribution. The position of the CM is given by \cite{goldstein}
\begin{equation}
\label{19}
x_{cm} = \frac{1}{M}\int x dm. 
\end{equation}
Note that $dm= \mu \lambda dx$, where $\mu$ is the mass of the particles, which gives us that
\begin{equation}
\label{20}
\int dm = \int_{0}^{\infty} \mu \lambda(x) dx = M,
\end{equation}
as it should be. Therefore, from Eq.~\eqref{19}, we have that the position of the CM becomes \cite{arfken}
\begin{equation}
\label{21}
x_{cm} = \frac {\mu} {M} \int_{0}^{\infty}  x \lambda(x) dx = \sqrt {\frac{2}{\pi}} \sigma.
\end{equation}
Since the external force is proportional to the linear density of particles along the $x$-axis and the temperature at the wall, as shown in Eq.~\eqref{16}, which are given respectively by $\lambda(0) = \sqrt{2} N/\sqrt{\pi} \sigma = 2 N/\pi x_{cm}$ and $T=T(0)$, we have that  
\begin{equation}
\label{22}
F^{(0)}_{ext} = \lambda(0) K_{B} T(0) = \frac{2 N k_{B} T(0)}{\pi x_{cm}}.
\end{equation}
This result also shows that the external force is inversely proportional to the distance between the wall and the CM of the gas.
 
\section*{APPENDIX C: EXTERNAL FORCE EXERTED BY A WALL ON AN IDEAL GAS WHEN THE CENTER OF MASS IS MOVING}
\renewcommand{\theequation}{B-\arabic{equation}}
\setcounter{equation}{0}

Here we demonstrate that the force exerted on the gas by a wall, besides the contribution necessary to confine the particles which depends on the inverse of the distance to the CM, as shown in Appendix B, it gains an extra contribution if the CM is moving perpendicular to the wall. As we shall see, this contribution is proportional to the velocity of the CM. In the previous case, in which the CM is at rest, if we assume that the masses of all constituents of the system are the same, $m_{i} = m$, from Eq.~(\ref{12}) we have that 

\begin{equation}
\label{23}
\mathbf{V} = \frac{m}{M} \sum_{i=1}^{N} \mathbf{v}_{i} = \frac{1}{N} \sum_{i=1}^{N} \mathbf{v}_{i} = 0.
\end{equation}
This is telling us that the average velocity of the constituents is zero, $\langle \mathbf{v} \rangle = (1/N) \sum_{i=1}^{N} \mathbf{v}_{i} = 0$, which is in accordance to what the Maxwell-Boltzmann distribution says for an ideal gas in equilibrium inside a container. 

Let us now consider the case in which the CM of the gas has a nonzero velocity $\mathbf{V'}$ pointing perpendicular to some container wall, e.g., along the $x$ direction with the form $\mathbf{V'} = v_{x} \mathbf{\hat{x}}$, where $\mathbf{\hat{x}}$ is the unit vector pointing in the positive $x$ direction. For this to happen, it is necessary that each constituent $i$ of the gas that we are analyzing acquire an extra velocity $\mathbf{w}_{i}$ when compared to the previous case. That is,         
\begin{equation}
\label{24}
\mathbf{V'} = \frac{m}{M} \sum_{i=1}^{N} (\mathbf{v}_{i} + \mathbf{w}_{i}) = \frac{1}{N} \sum_{i=1}^{N} \mathbf{w}_{i} = v_{x} \mathbf{\hat{x}},
\end{equation}
where in the last equality we used the result of Eq.~(\ref{23}). This expression tell us that the average of the extra velocities in this configuration is given by $\langle \mathbf{w} \rangle = (1/N) \sum_{i=1}^{N} \mathbf{w}_{i} = v_{x} \mathbf{\hat{x}}$, which is equal to the CM velocity.

It is natural now to pose the question: what is the influence of this extra velocity of the constituents, which accounts for the nonzero CM velocity, on the force exerted by the wall on the gas? We provide an answer to this question by examining the average of the extra momentum $\langle \Delta \mathbf{p} \rangle$ transferred by the wall to the constituents per unit time. Under this perspective, the extra force due to the movement of the CM of the gas is (see Ref. \cite{bert})
\begin{equation}
\label{25}
\mathbf{F}^{'}_{ext} = \frac{n \langle \Delta \mathbf{p} \rangle}{\Delta t} = -\frac{2 n m v_{x} \mathbf{\hat{x}}}{\Delta t},
\end{equation}
where $n$ is the number of collisions between the constituents and the wall that occur in a given time interval $\Delta t$. Note that, for $\mathbf{F}^{'}_{ext}$ to be recorded with appreciable statistics, this time interval must be much longer than the time interval between consecutive collisions. We put the factor $2$ only because we are considering elastic collisions, which is not a requirement. The negative sign indicates that the vectors $\langle \Delta \mathbf{p} \rangle$ and $v_{x} \mathbf{\hat{x}}$ point in opposite directions.  

The average number of collisions per unit time $n /\Delta t$ is a function of both the temperature $T$ and the number of constituents of the gas $N$. Therefore, the extra force due to the movement of the CM has a nonzero $x$ component with the form ${F}^{'}_{ext}=-g(T,N) \dot{x}$. This force acts as a damping for the dynamics of the CM, with a damping factor $g$. Nevertheless, it is important to observe that we are not imposing the existence of a friction force in this model. Such behavior emerges naturally from our Newtonian analysis of the problem. Indeed, no dissipative element has been included as an ingredient in the present derivation. Overall, we have that the contributions of the presence of the gas and the motion of the CM to the external force, taken together, assume the general form $F_{ext} = F^{(0)}_{ext} + {F}^{'}_{ext}$, along the $x$ direction, i.e., 
\begin{equation}
\label{26}
F_{ext} (x,\dot{x})= \frac{f(T,N)}{x}-g(T,N) \dot{x}. 
\end{equation}   
This relation justifies the dynamical model of the CM presented in Sec. IV for the case of the gas confined in the finite length reservoir, in which the number of collisions between the constituents and the walls $n$ does not vary considerably with time. Also in this scenario, we did not consider the $\dot{x}$-dependent term for the diffusion in the box with infinite length, Eq.~(\ref{9}), because in that case the number of collisions, and hence the damping factor $g$, decrease rapidly with time.

%%%%%%%%%%%%%%%%%%%%%%%%%%%%%%%%%%%%%%%%%%%%%%%%%%%%%%%%%%%


\begin{thebibliography}{7}
\bibitem{me} This fact does not represent a violation of the second law of thermodynamics because the act of reversing the velocities requires an external agent. Thus, the system cannot be isolated.
\end{thebibliography}
\end{document}